\documentclass[prb,floatfix,twocolumn,showpacs,amsmath,amssymb]{revtex4}
\usepackage{graphicx}
\usepackage{dcolumn}
\usepackage{bm}
\begin{document}

\title{Strong renormalization of the Fermi-surface topology close to the 
Mott transition}
\author{Luca F. Tocchio,$^{1}$ Federico Becca,$^{2}$
        and Claudius Gros$^{1}$ 
        }
\affiliation{
$^{1}$Institute for Theoretical Physics, 
       University of Frankfurt, 
       Max-von-Laue-Stra{\ss}e 1, D-60438 Frankfurt a.M., Germany \\
$^{2}$CNR-IOM-Democritos National Simulation Centre 
       and International School for Advanced Studies (SISSA), 
       Via Bonomea 265, I-34136, Trieste, Italy
            }

\date{\today} 

\begin{abstract}
The underlying Fermi surface is a key concept for strongly-interacting electron
models and has been introduced to generalize the usual notion of the Fermi 
surface to generic (superconducting or insulating) systems. By using improved
correlated wave functions that contain backflow and Jastrow terms, we examine 
the two-dimensional $t{-}t^\prime$ Hubbard model and find a non-trivial 
renormalization of the topology of the underlying Fermi surface close to the 
Mott insulator. Moreover, we observe a sharp crossover region, which arises 
from the metal-insulator transition, from a weakly interacting metal at small 
coupling to a resonating valence-bond superconductor at intermediate coupling. 
A violation of the Luttinger theorem is detected at low hole dopings. 
\end{abstract}


\maketitle

\section{Introduction}

The single-band Hubbard model with extended hopping on the square lattice
has been widely investigated since the appearance of the high-temperature 
superconductors. Indeed, it is believed to represent the minimal model that 
is necessary to describe the electronic correlations in the Copper-Oxygen 
planes of Cuprate materials. A very rich variety of phases has been discussed
in this context, including antiferromagnetism, superconductivity, 
charge-density waves, and non-Fermi-liquid metals.~\cite{carlson} 
Studies with various numerical techniques, ranging from dynamical mean-field 
theory (DMFT),~\cite{DMFT} including its cluster 
extensions,~\cite{jarrell,kancharla,civelli,jarrell2,jarrell3} to quantum Monte
Carlo techniques,~\cite{hirsch,white,paramekanti,sorella,varney} as well as 
analytic approaches,~\cite{rice,shastry} have addressed hotly debated 
topics like the nature of the pseudo-gap phase or the superconducting 
correlations. 

Landau~\cite{landau} and Luttinger~\cite{luttinger} have shown that the Fermi
surface, which is the locus of gapless electronic excitations in $k$-space,
represents a pivoting concept in the theory of Fermi liquids. 
The generalization of this notion for gapped systems (e.g., superconductors or
Mott insulators) leads to the idea that there is an underlying Fermi surface 
that becomes gapped because of some symmetry breaking (leading to 
superconductivity) or electronic correlation (leading to a Mott insulator).
In particular, the underlying Fermi surface can be defined by the locus of 
points where $\textrm{Re}\,G(k,\omega=0)$ changes sign ($G(k,\omega)$ is the 
single-particle Green's function), passing either to infinity 
(for usual Fermi liquids with well defined quasiparticles) or to 
zero.~\cite{dzyaloshinskii} The question of determining the underlying Fermi 
surface is of central importance in strongly correlated systems, particularly 
in view of experiments with angle-resolved photoemission spectroscopy 
(ARPES).~\cite{yang} From the theoretical point of view, there are few attempts
to study the topology of the Fermi surface in correlated systems; furthermore,
they are limited to mean-field approaches,~\cite{sensarma,grosPNAS,ruegg} 
including recent calculations with cluster DMFT.~\cite{civelli2}
Here, we make a substantial step forward and consider non-perturbative
calculations directly in a two-dimensional system. We use variational wave 
functions containing backflow correlations, which have been shown to be very 
accurate in the Hubbard model,~\cite{tocchiosquare,tocchioback} and focus our
attention on the topology of the underlying Fermi surface and the 
superconducting properties. We show that a strong renormalization of the 
underlying Fermi surface takes place close to half filling, when the Mott 
insulator is approached. In particular, the renormalization to perfect nesting
occurring at the interaction-driven metal-insulator transition opens a new 
perspective on the crucial role of momentum dependence in describing the Mott 
transition. 

At finite dopings we find a sharp crossover line, which emerges from the
Mott-Hubbard transition point at half filling and separates a weakly-correlated
metal and a strong-coupling superconducting state, in agreement with a recent 
observation of an unconventional metallic state at finite 
dopings.~\cite{fournier2010} Our findings could be also related to a recently 
proposed first-order line separating two metallic phases, one with a pseudogap
and one without.~\cite{sordi} For the superconducting order parameter we 
obtain a sizable signal for moderately large on-site interactions, i.e., 
$U/t \simeq 7 - 10$.

The paper is organized as follows: in section~\ref{sec:model}, we introduce 
the Hamiltonian and describe our variational wave function; in 
section~\ref{sec:results}, we present our numerical results and, finally, in 
section~\ref{sec:conc} we draw the conclusions.

\section{Model and variational wave function}\label{sec:model}

We consider the Hubbard model with extended hopping on a two-dimensional 
square lattice, 
\begin{equation}\label{eq:hamiltonian}
{\cal H} = -t \hspace{-0.5ex} \sum_{\langle ij\rangle\sigma} 
     \hspace{-1ex} c^{\dagger}_{i\sigma} c^{\phantom{\dagger}}_{j\sigma}
     -t^\prime \hspace{-0.5ex} \sum_{\langle\langle ij\rangle\rangle\sigma} 
     \hspace{-1ex} c^{\dagger}_{i\sigma} c^{\phantom{\dagger}}_{j\sigma}  
     +\textrm{h.c.} 
     +U \sum_{i} n_{i\uparrow} n_{i\downarrow},
\end{equation}
where $c^{\dagger}_{i\sigma}$ ($c^{\phantom{\dagger}}_{i\sigma}$) denotes the 
electron creation (destruction) operator of one electron on site $i$ with spin 
$\sigma=\uparrow,\downarrow$, $\langle ij\rangle$ and 
$\langle\langle ij\rangle\rangle$ indicate nearest and next-nearest neighbor 
sites respectively; 
$n_{i\sigma}=c^{\dagger}_{i\sigma}c^{\phantom{\dagger}}_{i\sigma}$ is the 
electron density; $t$ and $t^\prime$ are the nearest and next-nearest neighbor
hopping amplitudes, and $U$ is the on-site Coulomb repulsion. Calculations
are performed on 45-degree tilted clusters with $L=2\times l^2$ sites ($l$ 
being an odd integer) and periodic boundary conditions. The number of electrons
is $N$, such that the hole doping is $\delta=1-n$, with $n=N/L$.

\begin{figure}
\includegraphics[width=0.85\columnwidth]{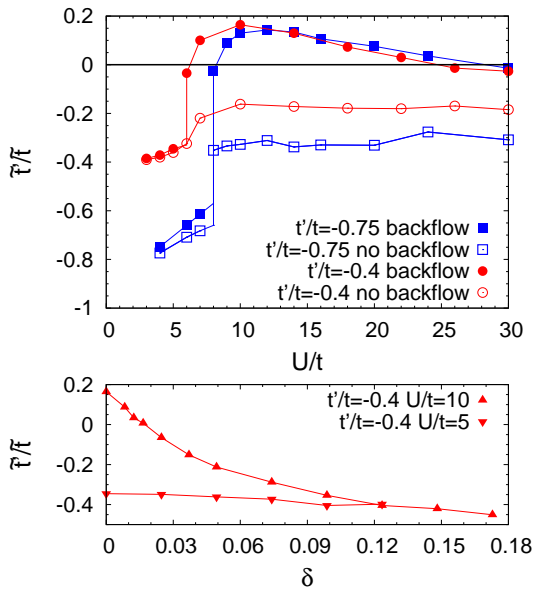}
\caption{\label{fig:ren}
(Color online) Upper panel: variational hopping ratio 
$\tilde{t}^\prime/\tilde{t}$ as a function of $U/t$ for $t^\prime/t=-0.4$ 
and $-0.75$. Full (empty) symbols refer to the presence (absence) 
of backflow correlations in the variational wave function. Lower panel: 
variational hopping ratio $\tilde{t}^\prime/\tilde{t}$ as a function of the
doping $\delta$ for $U/t=5$ and $10$ with $t^\prime/t=-0.4$. Here, only results
with backflow terms are shown. Data are shown for a $L=162$ lattice; in the 
lower panel few points on a $L=242$ lattice are added close to half filling.}
\end{figure}

The question of determining the underlying Fermi surface has been addressed
by using a renormalized mean-field approach in 
Refs.~\onlinecite{sensarma,grosPNAS}. Here, we include electronic correlations
in a non-perturbative way. In a first step, we construct uncorrelated wave 
functions given by the ground state $|\rm{BCS}\rangle$ of a superconducting 
Bardeen-Cooper-Schrieffer (BCS) Hamiltonian:~\cite{grosbcs,zhang}
\begin{equation}\label{eq:meanfield}
{\cal H}_{\rm{BCS}} = \sum_{k\sigma} \xi_k 
c^{\dagger}_{k\sigma} c^{\phantom{\dagger}}_{k\sigma}
+ \sum_{k} \Delta_k 
c^{\dagger}_{k\uparrow} c^{\dagger}_{-k\downarrow} + \rm{h.c.},
\end{equation}
where both the free-band dispersion $\xi_k$ and the pairing amplitudes 
$\Delta_k$ are variational functions. We use the parametrization
\begin{eqnarray}
\xi_k&=& -2\tilde{t}(\cos k_x +\cos k_y)
-4\tilde{t}^\prime\cos k_x \cos k_y -\mu \\
\Delta_k&=& \,2\Delta_{\textrm{BCS}}(\cos k_x -\cos k_y), 
\label{eq:effective} 
\end{eqnarray}
where the effective hopping amplitude $\tilde{t}^\prime$, the effective 
chemical potential $\mu$, and the local pairing field $\Delta_{\textrm{BCS}}$
are variational parameters to be optimized. The parameter $\tilde{t}$ is kept
fixed to set the energy scale. We also considered longer-range effective 
hopping parameters in Eq.~(\ref{eq:effective}) finding that all hoppings 
beyond the ones present in the Hamiltonian~(\ref{eq:hamiltonian}) are 
optimized to zero.

\begin{figure}
\includegraphics[width=1.0\columnwidth]{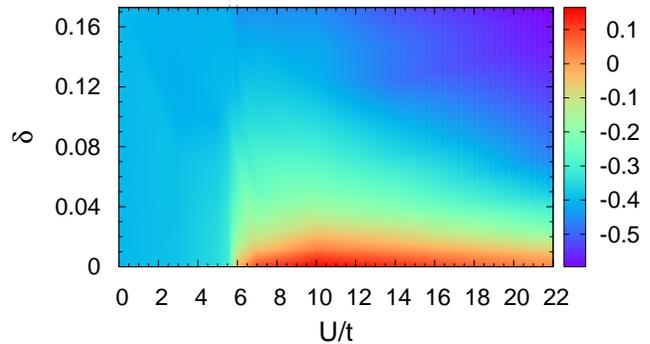}
\caption{\label{fig:ren_hopping}
(Color online) Variational hopping ratio $\tilde{t}^\prime/\tilde{t}$ as a 
function of doping $\delta$ and $U/t$ for the case with $t^\prime/t=-0.4$. 
The density plot is obtained by using results with $L=162$ and $242$.}
\end{figure}

The correlated state $|\Psi_{\textrm{BCS}}\rangle$, without backflow terms, 
is then given by $|\Psi_{\textrm{BCS}}\rangle = {\cal J} |\textrm{BCS}\rangle$, 
where ${\cal J}=\exp(-1/2 \sum_{ij} v_{ij} n_i n_j)$ is a density-density 
Jastrow factor (including the on-site Gutzwiller term $v_{ii}$), with the 
$v_{ij}$'s being optimized for every independent distance $|i-j|$. Notably, 
within this kind of wave function, it is possible to obtain a pure (i.e., 
non-magnetic) Mott insulator for a sufficiently singular Jastrow factor 
$v_q \sim 1/q^2$ ($v_q$ being the Fourier transform of $v_{ij}$), while a
superconducting (metallic) state is found whenever $v_{q}\sim 1/q$ and
$\Delta_{\textrm{BCS}}>0$ ($\Delta_{\textrm{BCS}}=0$).~\cite{capello}

A size-consistent and efficient way to further improve the correlated state
$|\Psi_{\rm{BCS}}\rangle$ for large on-site interactions is based on backflow 
correlations. In this approach, each orbital that defines the unprojected state
$|\textrm{BCS}\rangle$ is taken to depend upon the many-body configuration,
such to incorporate virtual hopping processes.~\cite{tocchiosquare,tocchioback}
All results presented here are obtained by fully incorporating the backflow 
corrections and optimizing individually every variational parameter in $\xi_k$
and $\Delta_k$, in the Jastrow factor ${\cal J}$, as well as for the backflow 
corrections.

\section{Results}\label{sec:results}
\subsection{Fermi-surface renormalization}

We start our analysis by discussing the evolution of the underlying 
Fermi surface topology as a function of hole doping $\delta$ and 
interaction strength $U/t$. Within the variational approach, the underlying 
Fermi surface can be easily defined and corresponds to the highest 
occupied momenta $\xi_k$. Indeed, $E_k=\sqrt{\xi_k^2+\Delta_k^2}$ 
corresponds, within renormalized mean-field theory,~\cite{zhang} 
to the excitation spectrum of projected Bogoliubov modes and 
hence $\xi_k$ to the dispersion of the renormalized quasiparticles. 
Here, we would like to stress that, although $\xi_k$ corresponds to the 
``non-interacting'' spectrum, it may be strongly renormalized because of the 
full optimization of the variational wave function: due to the presence of 
Jastrow and backflow terms, non-trivial values of $\tilde{t}^\prime$ and 
$\Delta_{\textrm{BCS}}$ may be obtained.

\begin{figure}
\includegraphics[width=0.85\columnwidth]{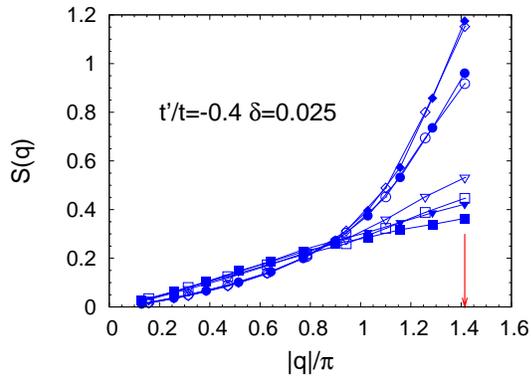}
\caption{\label{fig:S_q}
(Color online) Static spin-spin correlations $S(q)$ as a function of $|q|/\pi$
along the $\Gamma=(0,0)$ to $M=(\pi,\pi)$ direction, for $t^\prime/t=-0.4$. 
Data are presented for $U/t=5$ (squares), $U/t=5.6$ (triangles), $U/t=6$ 
(circles) and $U/t=7$ (diamonds). Full (empty) symbols refer to $L=242$
($L=162$). The location of the point $Q=(\pi,\pi)$ is marked with an arrow.}
\end{figure}

Firstly, we consider the half-filled case, where, by increasing the ratio
$U/t$ a metal-insulator transition is encountered. We present the results 
for $t^\prime/t=-0.4$ and $t^\prime/t=-0.75$, see Fig.~\ref{fig:ren}. 
In both cases, a renormalization towards perfect nesting, namely
$\tilde{t}^\prime/\tilde{t}=0$, is shown to occur both at the metal-insulator 
transition and in the limit $U/t \to \infty$. In the metallic phase,
the ratio $\tilde{t}^\prime/\tilde{t}$ is only slightly modified with respect
to its bare value, and this result is found both with and without backflow
correlations, demonstrating that even the simple wave function without backflow
may capture a correct description of the metallic phase. Then, at the
metal-insulator transition, the variational ratio of the hopping parameters 
is strongly renormalized in presence of backflow correlations, i.e.,
$\tilde{t}^\prime/\tilde{t} \to 0$, driving the underlying Fermi surface to
be perfectly nested. This fact suggests that a full momentum resolution is a 
crucial ingredient to properly describe a metal-insulator transition. 
Remarkably, this renormalization does not occur if backflow correlations are 
not included. By further increasing the on-site Coulomb repulsion, 
the renormalized hopping ratio changes sign and then decreases again
to zero at large $U/t$.~\cite{sign} This renormalization, which is again 
possible only in presence of backflow correlations, is in agreement with 
renormalized mean-field studies of the $t{-}J$ model,~\cite{grosPNAS} where
the Fermi surface renormalizes to perfect nesting for $\delta \to 0$, and with 
a slave-spin study of the Hubbard model in the large-interaction 
limit.~\cite{ruegg} Note that for $U/t>25$ the numerical accuracy for the 
ratio $\tilde{t}^\prime/\tilde{t}$ is not enough to distinguish a finite 
from a vanishing value, see Fig.~\ref{fig:ren}. 

\begin{figure}
\includegraphics[width=0.9\columnwidth]{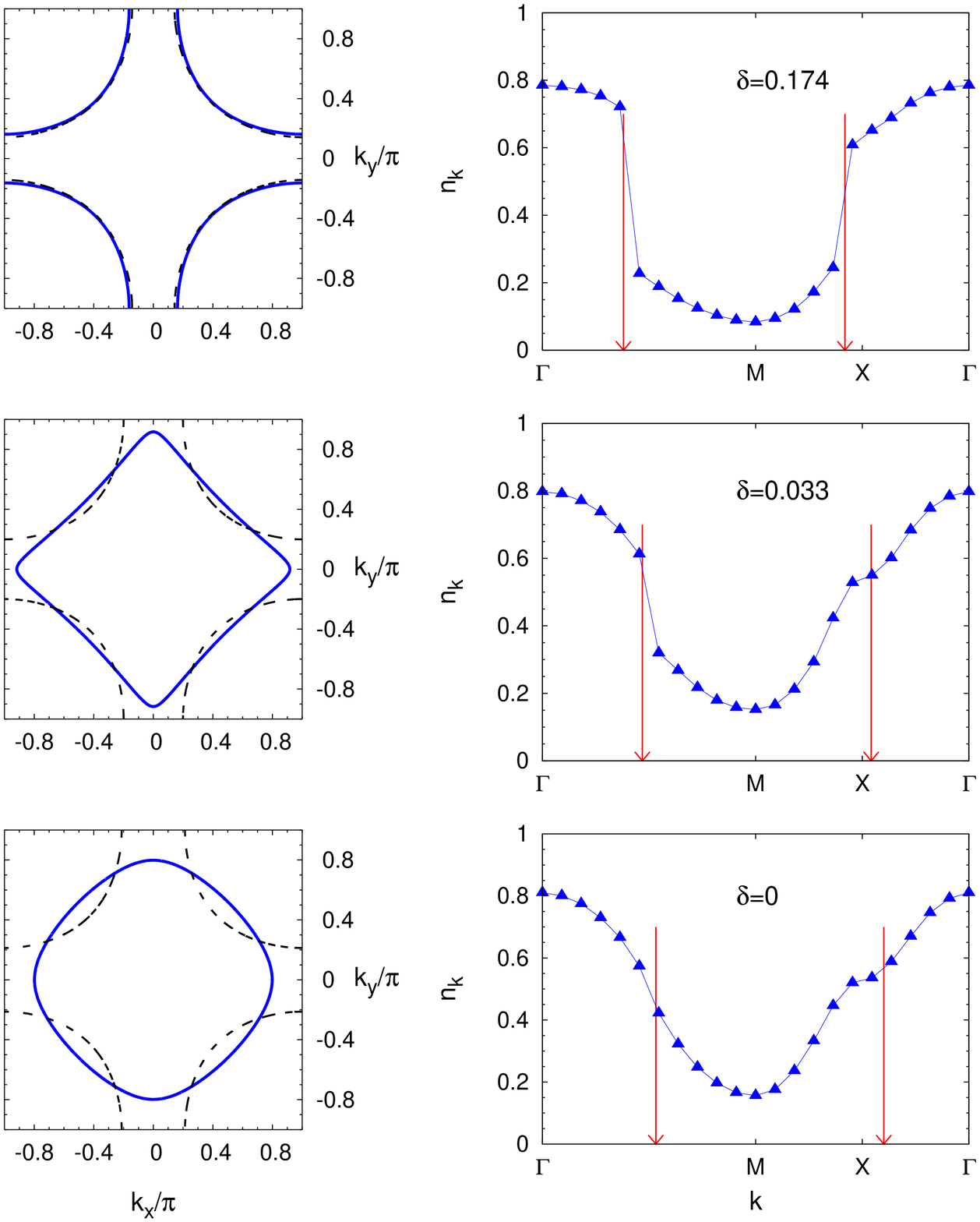}
\caption{\label{fig:n_k}
(Color online) Left panels: the underlying Fermi surface defined by $\xi_k=0$ 
(solid blue lines) and the non-interacting Fermi surface $\xi^0_k=0$
(dashed black lines) for the same parameter values of the corresponding right 
panels (see text for more details). Right panels: momentum distribution 
function $n_k$ along the path in the Brillouin zone connecting the points 
$\Gamma=(0,0)$, $M=(\pi,\pi)$, and $X=(\pi,0)$. Data are shown at $U/t=10$ and 
$t^\prime/t=-0.4$ for a $L=242$ lattice size at three dopings ($\delta=0$, 
$0.033$, and $0.174$). Arrows indicate the position of the underlying Fermi 
surface, as found in left panels.}
\end{figure}

We would like to mention that a Fermi-surface renormalization close to the 
Mott transition has already been described in the one-dimensional Hubbard 
model.~\cite{tocchio1D} Also in that case, a perfectly nested Fermi surface, 
i.e., $k_F=\pm \pi/2$, is found. This consideration enforces the idea that the
variational hopping ratio in two dimensions does not vanish accidentally at 
the Mott transition, but is renormalized in order to get perfect nesting of 
the underlying Fermi surface. 

The variational hopping changes very rapidly by doping the Mott insulator and 
tends towards the bare value at high values of hole dopings, see the case 
$U/t=10$ in  Fig.~\ref{fig:ren}. In summary, we present in 
Fig.~\ref{fig:ren_hopping} a complete density plot for the variational hopping
$\tilde{t}^\prime/\tilde{t}$ as a function of doping $\delta$ and $U/t$, with 
bare $t^\prime/t=-0.4$. While we have almost no renormalization when the 
half-filled case is metallic (i.e., for $U/t \lesssim 6$), we observe that a 
strong renormalization close to half filling is present for $U/t \gtrsim 6$. 
Remarkably, the metal-insulator transition occurring at half filling comes with
a sharp crossover line at finite doping that we can characterize through an 
abrupt change in the value of $\tilde{t}^\prime/\tilde{t}$. This crossover 
separates a weakly-correlated metal at low $U/t$ and a strongly renormalized 
state at intermediate/strong values of $U/t$. This result is in agreement with
a recent ARPES study on YBa$_2$Cu$_3$O$_{6+x}$, which found indications for an 
unconventional metallic state.~\cite{fournier2010} 

\subsection{Magnetic correlations}

We provide here further evidence of the sharp crossover line by calculating 
the static spin-spin correlations, defined as 
\begin{equation}
S(q) = \frac{1}{L} \sum_{m,n} e^{i q (R_m-R_n)} \langle S_m^z S_n^z \rangle,
\end{equation}
where $S_m^z$ is the $z$-component of the spin operator on site $m$.
The presence of (short-range) antiferromagnetic correlations is signaled by 
the appearance of a (non-diverging) peak in $S(q)$, located at $Q=(\pi,\pi)$.
As shown in Fig.~\ref{fig:S_q} for the doping $\delta=0.025$, the correlated 
resonating-valence bond (RVB) state at $U/t \gtrsim 6$ is characterized by 
antiferromagnetic correlations that are strongly enhanced with respect to the 
metallic phase at $U/t \lesssim 6$. The two regimes are clearly separated 
by a jump in the value of $S(Q)$; the short-range nature of the 
antiferromagnetic correlations is confirmed by a size scaling study.

\begin{figure}
\includegraphics[width=1.0\columnwidth]{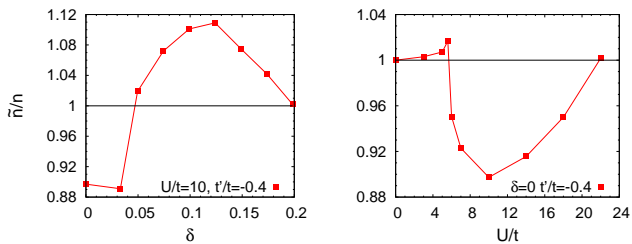}
\caption{\label{fig:Luttinger}
(Color online) Left panel: Ratio between the number of electrons enclosed by 
the renormalized Fermi surface, $\tilde{n}$, and the number of electrons 
effectively present in the system, $n$, as a function of doping. Data refer 
to the case $U/t=10$ and $L=242$. Right panel: Same quantity as in the left 
panel as a function of $U/t$ at half filling, for $L=242$.}
\end{figure}

\subsection{Momentum distribution function} 

In Fig.~\ref{fig:n_k}, we compare the results of the underlying Fermi surface 
with the ones for the momentum distribution
$n_k=\langle c_{k\sigma}^{\dagger} c_{k\sigma}^{\phantom{\dagger}}\rangle$,
for $U/t=10$ and three different dopings. 
Interestingly, in the low-doping regime, the area enclosed by $\xi_k=0$ is 
different from the one enclosed by the non-interacting system $\xi^0_k=0$, 
where $\xi^0_k= -2t(\cos k_x+\cos k_y) -4t^\prime\cos k_x \cos k_y -\mu_0$,
$\mu_0$ being the bare chemical potential. At half filling, the underlying 
Fermi surface is closed, but the system is insulating and consistently $n_k$ 
shows a completely smooth behavior. This is due to the presence of a strong 
Jastrow factor $v_q \sim 1/q^2$ which is able to remove the singularities 
present in $|\textrm{BCS}\rangle$.~\cite{capello2} As soon as a small doping 
is considered, the system becomes conducting and a finite jump in $n_k$ 
appears along the nodal direction $\Gamma \to M$, where the pairing amplitude
$\Delta_k$ vanishes. At small dopings, the variational hopping 
$\tilde{t}^\prime/\tilde{t}$ undergoes an abrupt change (see 
Fig.~\ref{fig:ren_hopping}) and, therefore, the Fermi surface becomes rapidly
open, see Fig.~\ref{fig:n_k}. Finally, at large dopings, the superconducting 
gap vanishes and a full Fermi surface is recovered also in the momentum 
distribution, i.e., a finite jump is also detected along the $M \to X$ 
direction. Additional results on the evolution of the underlying Fermi surface
as a function of doping for $U/t=10$ and $U/t=5$ can be found in the
Appendix~\ref{sec:Appendix}.

In the large doping case, the jump along the $M \to X$ direction coincides with
the position of the underlying Fermi surface. When the doping is small and 
there is no jump along the $M\to X$ direction, one could extend the previous 
concept and associate the location of the maximal gradient of $n_k$ with the 
position of the underlying Fermi surface. However, as shown in 
Fig.~\ref{fig:n_k}, this approach would lead to incorrect results, since the 
underlying Fermi surface is closer to the $X$ point with respect to the 
location of the maximal gradient. Therefore, our results suggest that, when 
doping is small and electronic correlation is important, some caution should 
be taken in deriving the Luttinger surface from the location of the maximal 
gradient of $n_k$,~\cite{randeria,mesot} similarly to what has been discussed
in Ref.~\onlinecite{grosPNAS} for the $t{-}J$ model.

\begin{figure}
\includegraphics[width=0.85\columnwidth]{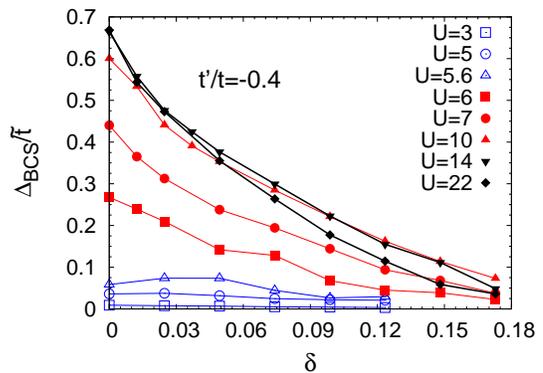}
\caption{\label{fig:Delta_BCS}
(Color online) Pairing amplitude $\Delta_{\textrm{BCS}}/\tilde{t}$ as a 
function of doping for $t^\prime/t=-0.4$ and different values of $U/t$. 
Data are presented for a $L=162$ lattice size.}
\end{figure}

\begin{figure*}
\includegraphics[width=0.9\textwidth]{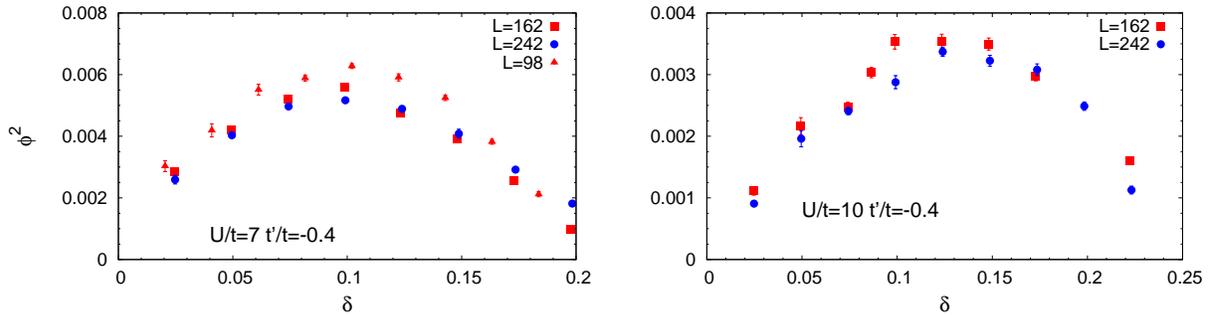}
\caption{\label{fig:order_parameter}
(Color online) Superconducting order parameter squared $\phi^2$ as a function 
of doping for $t^\prime/t=-0.4$ and $U/t=7$ (left panel) and $U/t=10$ (right 
panel).}
\end{figure*}

\subsection{Luttinger sum rule}

The renormalization of the underlying Fermi surface leads to a violation 
of the Luttinger sum rule. Let us denote by $\tilde{n}$ the number of electrons
that are enclosed by the renormalized Fermi surface $\xi_k=0$. As shown in 
Fig.~\ref{fig:n_k}, close to half filling, the $\xi_k=0$ contour is 
electron-like, containing less electrons than $n$, i.e., $\tilde{n}/n<1$,
while for $0.05 \lesssim \delta \lesssim 0.2$, the contour is hole-like and 
$\tilde{n}/n>1$. For larger values of the doping, the Luttinger count holds.
The summary for the Luttinger sum rule is reported in Fig.~\ref{fig:Luttinger}. 
Our results are slightly different from the ones found at the simple 
mean-field level,~\cite{sensarma} especially close to the Mott insulator where
we find that $\delta n \simeq -0.1$. In Fig.~\ref{fig:Luttinger}, we also 
present the violation of the Luttinger sum rule at half filling, as a function
of $U/t$. In the insulating state ($U/t \gtrsim 6$) the underlying Fermi 
surface defined by $\xi_k=0$ encloses less electrons than $n$. This result is 
compatible with the electron-like nature of the contour, due to the strong 
renormalization of the variational hopping parameters. On the contrary, in the
metallic region we only observe a tiny violation, in particular close to the 
metal-insulator transition. This tiny violation can be related to the amount 
of correlation present in the low-$U$ metallic state and it is indeed 
significantly smaller than the degree of correlation that can be observed in 
the RVB state emerging from the Mott insulator upon doping, see 
Fig~\ref{fig:Luttinger}.

\subsection{Pairing amplitude and pair-pair correlations}

We discuss now the variational pairing amplitude $\Delta_{\textrm{BCS}}$, 
which measures the tendency to create resonating singlets, and the actual 
pair-pair correlations
$\langle\Delta(r)\rangle=\langle S^{\phantom{\dagger}}_{r}S^{\dagger}_0\rangle$,
where $S^{\dagger}_r= c^{\dagger}_{r\uparrow}c^{\dagger}_{r+x\downarrow}
-c^{\dagger}_{r\downarrow}c^{\dagger}_{r+x\uparrow}$.
At weak-coupling (when a conducting state is found at half filling), 
$\Delta_{\textrm{BCS}}$ remains relatively small and does not change much with
doping, see Fig~\ref{fig:Delta_BCS}. For $U/t \gtrsim 6$ (when an insulating 
state is found at half filling), $\Delta_{\textrm{BCS}}$ has a sizable value 
and decreases monotonically as a function of the doping. This behavior of the 
pairing amplitude has been related to the pseudo-gap of the normal phase of 
Cuprate materials.~\cite{paramekanti} 

In Fig~\ref{fig:order_parameter}, we report the results for the superconducting
order parameter $\phi^2=\lim_{r\to \infty}\Delta(r)$, similarly to what has 
been done in previous calculations for the Hubbard and $t-J$ 
models.~\cite{paramekanti,spanu} Our results are presented for $U/t=7$ and 
$10$, where superconductivity develops at finite doping. 
For smaller values of the on-site interactions, a much smaller signal is 
obtained. By increasing $U/t$, the optimal doping becomes larger 
and the magnitude of the order parameter decreases, suggesting that 
intermediate values of $U/t$ are optimal for maximal critical 
temperatures $T_c$, when assuming that $T_c$ scales with $\phi$.

\section{Conclusions}\label{sec:conc}

Our calculations represent a first attempt to trace the Fermi surface
renormalization in a truly two-dimensional system where the electron-electron
correlation is treated beyond simple mean-field approaches. A substantial 
deviation from the Luttinger sum rule is observed for large values of $U/t$, 
where the topology of the underlying Fermi surface changes from electron-like
to hole-like by increasing the doping. It would be very interesting to verify
these results on correlated materials, like Cuprate superconductors, where
ARPES probes should be able to detect changes in the topology of the
Fermi surface.

Furthermore, we show evidence of a sharp crossover region that 
originates from the metal-insulator transition and separates a 
weakly-correlated metal from a more correlated RVB superconductor at low 
dopings.

L.F.T. and C.G. acknowledge the support of the German Science Foundation 
through the Transregio 49.

\appendix
\section{Underlying Fermi surface}\label{sec:Appendix}

\begin{figure*}
\includegraphics[width=0.85\textwidth]{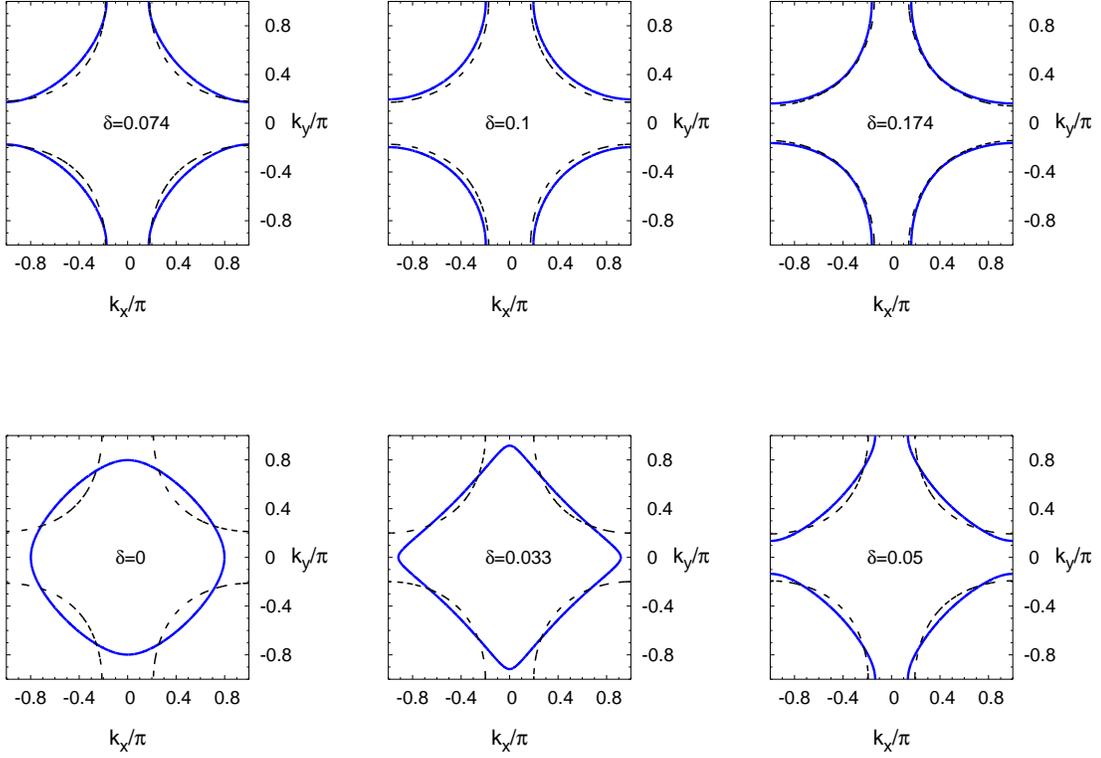}
\caption{\label{fig:Fermi_surface}
(Color online) The underlying Fermi surface defined by $\xi_k=0$ (solid blue 
lines) and the non-interacting Fermi surface $\xi^0_k=0$ (dashed black lines) 
as a function of doping. Data are presented at $U/t=10$ for $L=242$.}
\end{figure*}

In Fig.~\ref{fig:Fermi_surface}, we present a systematic plot of the underlying
Fermi surface defined by the contour $\xi_k=0$ and of the non-interacting 
Fermi surface, $\xi^0_k=0$, as a function of doping for the case $U/t=10$. 
The underlying Fermi surface evolves from electron-like, when 
$\delta \lesssim 0.05$, to hole-like, for $\delta \gtrsim 0.05$. When 
$\delta \sim  0.2$ the underlying Fermi surface becomes very close to the 
non-interacting one, as expected when doping becomes large and electronic 
correlation consequently less important.

\begin{figure*}
\includegraphics[width=0.85\textwidth]{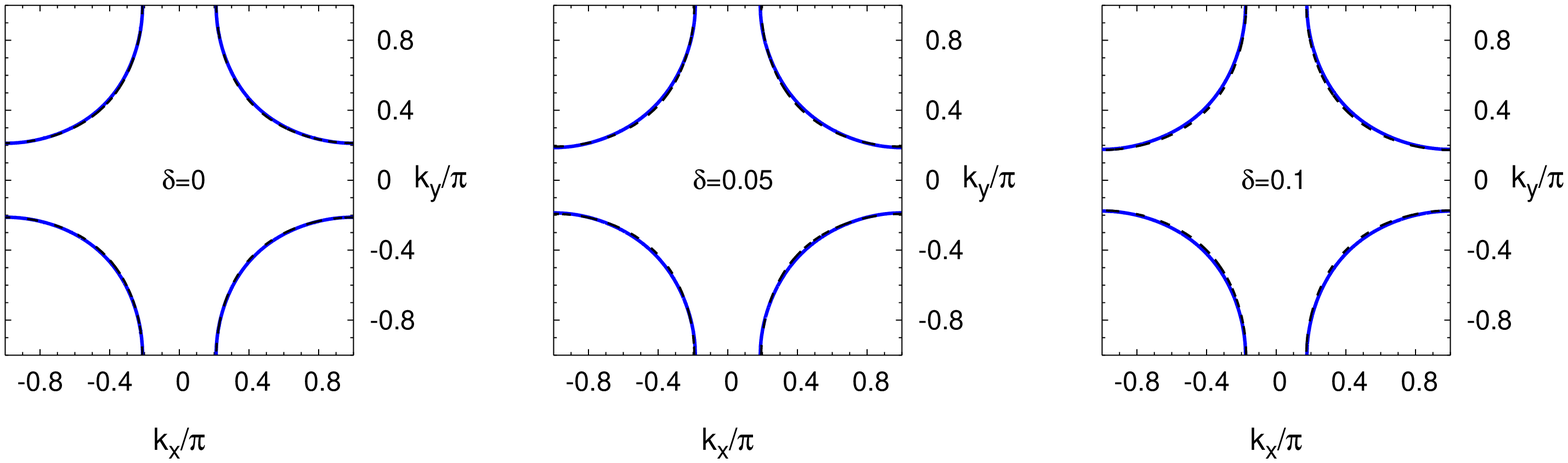}
\caption{\label{fig:Fermi_surface_U5}
(Color online) The underlying Fermi surface defined by $\xi_k=0$ 
(solid blue lines) and the non-interacting Fermi surface $\xi^0_k=0$
(dashed black lines) as a function of doping. Data are presented at $U/t=5$ 
for $L=242$.}
\end{figure*}

In Fig.~\ref{fig:Fermi_surface_U5}, we present also a plot of the
underlying Fermi surface as a function of doping for $U/t=5$. In this regime, 
the half-filled case is metallic and the renormalization of the variational 
parameters is very weak. Consequently, the underlying Fermi surface does not 
show a remarkable evolution as a function of doping and almost coincides with 
the non-interacting contour.   

\begin{thebibliography}{99}

\bibitem{carlson} See for example, E.W. Carlson, V.J. Emery, S.A. Kivelson, 
D. Orgad, {\it Concepts in high-temperature superconductivity} in 
``The Physics of Conventional and Unconventional Superconductors'' ed. by 
K.H. Bennemann and J.B. Ketterson (Springer-Verlag, 2003).

\bibitem{DMFT} A. Georges, G. Kotliar, W. Krauth, and M.J. Rozenberg, \rmp
{\bf 68}, 13 (1996).

\bibitem{jarrell}  T.A. Maier, M. Jarrell, T.C. Schulthess, P.R.C. Kent, 
and J.B. White, \prl {\bf 95}, 237001 (2005).

\bibitem{kancharla} S.S. Kancharla, B. Kyung, D. Senechal, M. Civelli, 
M. Capone, G. Kotliar, and A.-M.S. Tremblay, \prb {\bf 77}, 184516 (2008). 

\bibitem{civelli} M. Civelli, \prb {\bf 79}, 195113 (2009). 

\bibitem{jarrell2} K.-S. Chen, S. Pathak, S.-X. Yang, S.-Q. Su, D. Galanakis,
K. Mikelsons, M. Jarrell, and J. Moreno, \prb {\bf 84}, 245107 (2011).

\bibitem{jarrell3} S.-X. Yang, H. Fotso, S.-Q. Su, D. Galanakis, E. Khatami, 
J.-H. She, J. Moreno, J. Zaanen, and M. Jarrell, \prl {\bf 106}, 047004 (2011).

\bibitem{hirsch} J.E. Hirsch, \prb {\bf 31}, 4403 (1985).

\bibitem{white} S.R. White, D.J. Scalapino, R.L. Sugar, E.Y. Loh, 
J.E. Gubernatis, and R.T. Scalettar, \prb {\bf 40}, 506 (1989).

\bibitem{paramekanti} A. Paramekanti, M. Randeria, and N. Trivedi, \prl 
{\bf 87}, 217002 (2001).

\bibitem{sorella} S. Sorella, G.B. Martins, F. Becca, C. Gazza, L. Capriotti, 
A. Parola, and E. Dagotto, \prl {\bf 88}, 117002 (2002).
 
\bibitem{varney} C.N. Varney, C.-R. Lee, Z.J. Bai, S. Chiesa, M. Jarrell,
and R.T. Scalettar, \prb {\bf 80}, 075116 (2009).

\bibitem{rice} K.-Y. Yang, T.M. Rice, and F.-C. Zhang, \prb {\bf 73}, 174501 
(2006).

\bibitem{shastry} B.S. Shastry, \prl {\bf 107}, 056403 (2011); see also 
G.H. Gweon, B.S. Shastry, and G.D. Gu, \prl {\bf 107} 056404 (2011).

\bibitem{landau} L.D. Landau, JETP {\bf 3}, 920 (1957).

\bibitem{luttinger} J.M. Luttinger, Phys. Rev. {\bf 119}, 1153 (1960).

\bibitem{dzyaloshinskii} I. Dzyaloshinskii, \prb {\bf 68}, 085113 (2003).

\bibitem{yang} H.B. Yang, J.D. Rameau, Z.-H. Pan, G.D. Gu, P.D. Johnson,
H. Claus, D.G. Hinks, and T.E. Kidd, \prl {\bf 107}, 047003 (2011).

\bibitem{sensarma} R. Sensarma, M. Randeria, and N. Trivedi, \prl {\bf 98},
027004 (2007).

\bibitem{grosPNAS} C. Gros, B. Edegger, V.N. Muthukumar, and P.W. Anderson,
PNAS {\bf 103}, 14298 (2006).

\bibitem{ruegg} A. R\"uegg, S.D. Huber, M. Sigrist, \prb {\bf 81}, 155118 
(2010).

\bibitem{civelli2} M. Civelli, M. Capone, S.S. Kancharla, O. Parcollet, and 
G. Kotliar, \prl {\bf 95}, 106402 (2005).

\bibitem{tocchiosquare} L.F. Tocchio, F. Becca, A. Parola, and S. Sorella, 
\prb {\bf 78}, 041101(R) (2008); see also, F. Becca, L.F. Tocchio, and 
S. Sorella, J. Phys.: Conf. Ser. {\bf 145}, 012016 (2009).

\bibitem{tocchioback} L.F. Tocchio, F. Becca, and C. Gros, \prb {\bf 83}, 
195138 (2011).

\bibitem{fournier2010} D. Fournier, G. Levy, Y. Pennec, J.L. McChesney,
A. Bostwick, E. Rotenberg, R. Liang, W. N. Hardy, D.A. Bonn, I. S. Elfimov,
and A. Damascelli, Nature Physics {\bf 6}, 905 (2010).

\bibitem{sordi} G. Sordi, P. S\'emon, K. Haule, and A.-M. S. Tremblay, 
arXiv:1110.1392 (2011).

\bibitem{grosbcs} C. Gros, \prb {\bf 38}, 931(R) (1988).

\bibitem{zhang} F.C. Zhang, C. Gros, T.M. Rice, and H. Shiba, 
Supercond. Sci. Technol. {\bf 1}, 36 (1988).

\bibitem{capello} M. Capello, F. Becca, M. Fabrizio, S. Sorella, 
and E. Tosatti, \prl {\bf 94}, 026406 (2005).

\bibitem{sign} At half filling, the sign of $t^\prime$ (and, consistently
$\tilde{t}^\prime$) is irrelevant. We take the same sign (negative) that is 
used at finite dopings.

\bibitem{tocchio1D} L.F. Tocchio, F. Becca, and C. Gros, \prb {\bf 81}, 205109
(2010).

\bibitem{capello2} M. Capello, F. Becca, S. Yunoki, and S. Sorella, \prb 
{\bf 73}, 245116 (2006).

\bibitem{randeria} M. Randeria, H. Ding, J-C. Campuzano, A. Bellman, 
G. Jennings, T. Yokoya, T. Takahashi, H. Katayama-Yoshida, T. Mochiku, 
and K. Kadowaki, \prl {\bf 74}, 4951 (1995).

\bibitem{mesot} J. Mesot, M. Randeria, M.R. Norman, A. Kaminski, H.M. Fretwell,
J.C. Campuzano, H. Ding, T. Takeuchi, T. Sato, T. Yokoya, T. Takahashi, 
I. Chong, T. Terashima, M. Takano, T. Mochiku, and K. Kadowaki, \prb {\bf 63},
224516 (2001).
 
\bibitem{spanu} L. Spanu, M. Lugas, F. Becca, and S. Sorella, \prb {\bf 77}, 
024510 (2008).

\end{thebibliography}
\end{document}